\documentclass[manuscript]{acmart}

\usepackage{enumitem}
\usepackage{tikz}
\usepackage{array}
\usepackage{wrapfig}

\usetikzlibrary{external, shapes.geometric, arrows}
\usepackage{xspace}
\acmDOI{}          
\acmISBN{}

\AtBeginDocument{%
  \providecommand\BibTeX{{%
    \normalfont B\kern-0.5em{\scshape i\kern-0.25em b}\kern-0.8em\TeX}}}

\begin{document}

\tikzstyle{5_box_node} = [
    rectangle,
    rounded corners, 
    minimum width=2cm, 
    minimum height=1cm,
    text centered,
    text width = 2.5cm,
    draw=black,
]
\tikzstyle{3_box_node} = [
    rectangle,
    rounded corners, 
    minimum width=3cm, 
    minimum height=1cm,
    text centered,
    text width = 4cm,
    draw=black,
]
\tikzstyle{4_box_node} = [
    rectangle,
    rounded corners, 
    minimum width=3cm, 
    minimum height=1cm,
    text centered,
    text width = 3.2cm,
    draw=black,
]
\tikzstyle{arrow} = [thick,->,>=stealth]


\newcommand{\yaqing}[1]{\textcolor{violet}{#1}}
\newcommand{\vm}[1]{\textcolor{blue}{#1}}
\newcommand{\sssec}[1]{\vspace*{0.05in}\noindent\textbf{#1}}

\newcommand{\revision}[1]{\textcolor{black}{#1}}
\newcommand{\sys}{\text{FlexMind}\xspace}

\title{Scaffolding Flexible Ideation Workflows with AI in Creative Problem-Solving}

\author{Yaqing Yang}
\affiliation{%
  \institution{Carnegie Mellon University}
  \city{Pittsburgh, PA}
  \country{USA}}
\email{yaqingyy@cs.cmu.edu}

\author{Vikram Mohanty}
\orcid{0000-0001-6296-3134}
\affiliation{%
  \institution{Carnegie Mellon University}
  \city{Pittsburgh, PA}
  \country{USA}}
\email{vikrammohanty@acm.org}


\author{Nikolas Martelaro}
\affiliation{%
  \institution{Carnegie Mellon University}
  \city{Pittsburgh, PA}
  \country{USA}}
\email{nikmart@cmu.edu}

\author{Aniket Kittur}
\affiliation{%
  \institution{Carnegie Mellon University}
  \city{Pittsburgh, PA}
  \country{USA}}
\email{nkittur@cs.cmu.edu}

\author{Yan-Ying Chen}
\affiliation{%
  \institution{Toyota Research Institute}
  \city{Los Altos, CA}
  \country{USA}}
\email{yan-ying.chen@tri.global}

\author{Matthew K. Hong}
\affiliation{%
  \institution{Toyota Research Institute}
  \city{Los Altos, CA}
  \country{USA}}
\email{matt.hong@tri.global}

\renewcommand{\shortauthors}{Yang et al.}

\begin{abstract}
Divergent thinking in the ideation stage of creative problem‑solving demands that individuals explore a broad design space. Yet this exploration rarely follows a neat, linear sequence; problem‑solvers constantly shift among searching, creating, and evaluating ideas. Existing interfaces either impose rigid, step‑by‑step workflows or permit unguided free‑form exploration. To strike a balance between flexibility and guidance for augmenting people's efficiency and creativity, we introduce a human-AI collaborative workflow that supports a fluid ideation process. The system surfaces three opt-in aids: (1) high-level schemas to uncover alternative ideas, (2) risk analysis with mitigation suggestions, 
and (3) steering system-generated suggestions. Users can invoke these supports at any moment, allowing seamless back‑and‑forth movement among design actions to maintain creative momentum.

\end{abstract}



\maketitle

\section{Introduction}
\label{intro} 
Ideation is a critical, iterative phase of creative problem-solving where creators explore their solution space broadly. Rather than a one-shot brainstorm, it involves developing and refining both the problem formulation and solution ideas with constant information interchange~\cite{dorst2001creativity}. This process requires solvers to loop through searching, generating, and evaluating ideas while seeking information assistance to find inspiration and understand relevant knowledge~\cite{sawyer2003creativity}. Users often rely on web search~\cite{xu2024idea} and generative AI~\cite{kwon2023understanding} for the supplementary information.  However, these systems primarily serve as information sources rather than offering meaningful interactions that support users' cognitive flow. 
In the repetitive ideation cycle, people's actions occur in non-linear and highly individualized sequences~\cite{lee2024impact,gonccalves2021life}.
 Providing users with completely free-form tools, such as generic search engines without guidance, creates flexibility but risks information overload. Lack of constraints triggers the paradox of choice~\cite{schwartz2015paradox}, which can undermine creativity and efficiency during ideation~\cite{joyce2009blank,chua2008creativity,xu2024idea}.  Conversely, interfaces that enforce rigid sequences, such as tools with predefined action orders ~\cite{zhong2024ai} or linear chat-based systems~\cite{masson2024directgpt}, disrupt people's own thinking and creation processes. Therefore, effective creativity support must strike a middle ground: offering the right amount of guidance to nudge people to keep their creative momentum high while preserving people's own thinking flow.

\begin{figure*}[htb]
  \centering
  \includegraphics[width=\linewidth]{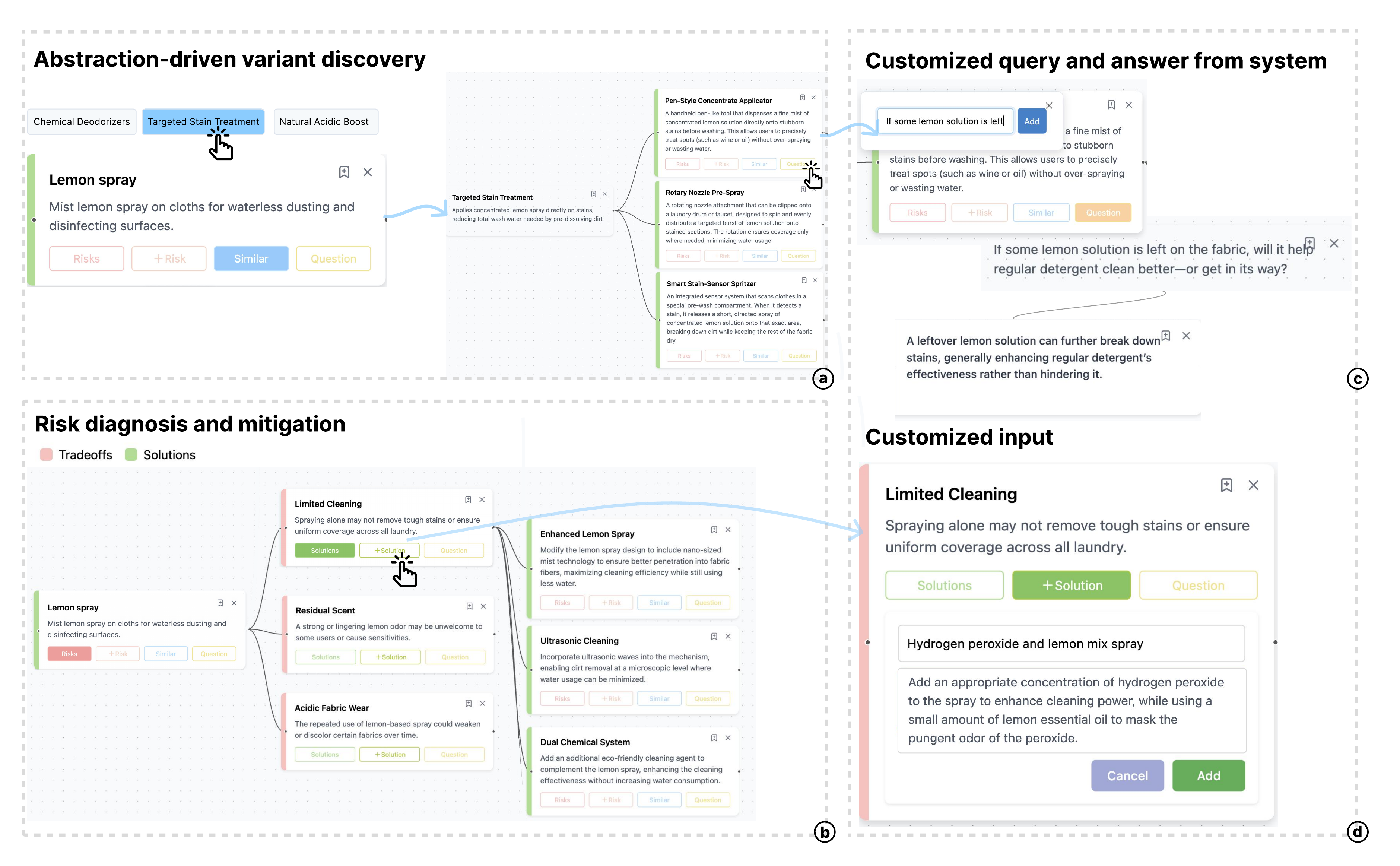}
  \caption{ The opt-in supports for flexible ideation workflow. a.Use abstraction to present the idea’s essential attributes and retrieves diverse implementations of them to provide similar ideas. b. Provide  feedback on risks and possible workarounds. c\&d. Allow users to steer the ideas received from the system. }
  \Description{The opt-in supports for flexible ideation workflow. a.Use abstraction to present the idea’s essential attributes and retrieves diverse implementations of them to provide similar ideas. b. Provide  feedback on risks and possible workarounds. c\&d. Allow users to steer the ideas received from the system. }
  \label{fig:components}
\end{figure*}

With extensive internal knowledge and contextual reasoning capability, large language models (LLMs) show promise as on-demand `thinking partners' capable of supporting flexible information needs~\cite{heyman2024supermind,shen2025ideationweb}. While previous work has explored AI-assisted ideation for different contexts~\cite{zhong2024ai,kang2025biospark,shen2025ideationweb}, we argue that what remains lacking is a set of generalizable and intuitive interactions that users can invoke as needed to foster effective collaboration with AI. Our goal is to augment rather than orchestrate people's thinking, respecting the non-linear nature of real-world ideation.  

Drawing on both prior research~\cite{son2024demystifying,fischer2018identifying,riche2025ai,xu2024idea} and our preliminary tests, we focus on addressing users’ three key information needs: discovering conceptually related ideas, critically assessing risks, and steering system-generated suggestions in response to personal insights. We propose a human-AI collaborative workflow offering three opt-in supports:

\begin{itemize}
    \item \textbf{Abstraction‑driven variant discovery .} When users are drawn to an idea, they often seek related variants but struggle to articulate what to search for~\cite{kwon2023understanding}. In open-ended tasks, reasoning is often tacit~\cite{son2024demystifying}, and the underlying concepts driving interest may be unclear. For example, suppose a designer exploring the task `cleaning laundry with less water' finds the idea `lemon spray' intriguing. In that case, they may be interested in concepts like targeted application or natural acids. Our system abstracts key attributes from the selected idea and retrieves diverse implementations of those attributes, balancing conceptual relevance with variety (Figure~\ref{fig:components}a)~\cite{root2001sparks}.
    \item \textbf{Risk diagnosis and mitigation.} Risk analysis expands the design space by identifying potential drawbacks and surfacing alternative approaches~\cite{fischer2018identifying}. Our system provides such feedback to help users judge whether to pivot, iterate, or double down. For instance, when evaluating using `lemon spray' for laundry, the system identifies risks like `limited cleaning' and suggests mitigations, such as improving the mist technology for better penetration into fabrics(Figure~\ref{fig:components}b). 
    \item \textbf{Idea steering.} Effective co-creation requires allowing users to steer system-generated ideas~\cite{riche2025ai}. Therefore, in our workflow, users can contribute their own thoughts, which the system uses to tailor its responses (Figure~\ref{fig:components}c). Users can also pose ad-hoc questions, receiving concise, context-aware answers that keep creative momentum high(Figure~\ref{fig:components}d)..
\end{itemize}

By integrating these supports into a human–AI collaborative workflow, users can quickly access the information they need without disrupting their thinking. The supports also break down reasoning into traceable units, making the process more explicit. This enables users to revisit earlier stages, preserving the flexible, nonlinear exploration essential to divergent thinking and amplifying designers’ natural creative loops.



\section{System Walkthrough}

\begin{figure}[t]
  \centering
  \includegraphics[width=\linewidth]{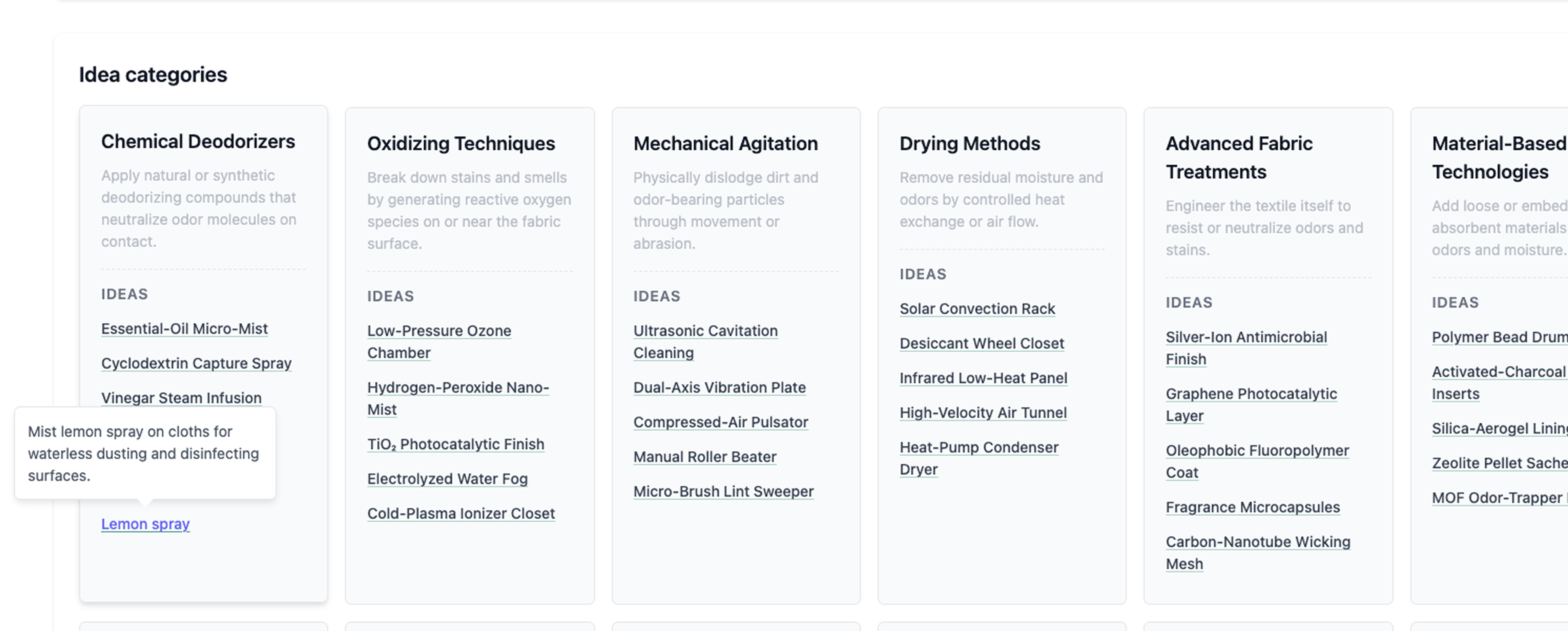}
  \caption{AI initializes the design space with idea categories and ideas for users to view.}
  \Description{AI initializes the design space with idea categories and ideas for users to view.}
  \label{fig:overview-b}
\end{figure}

\begin{figure}[t]
  \centering
  \includegraphics[width=\linewidth]{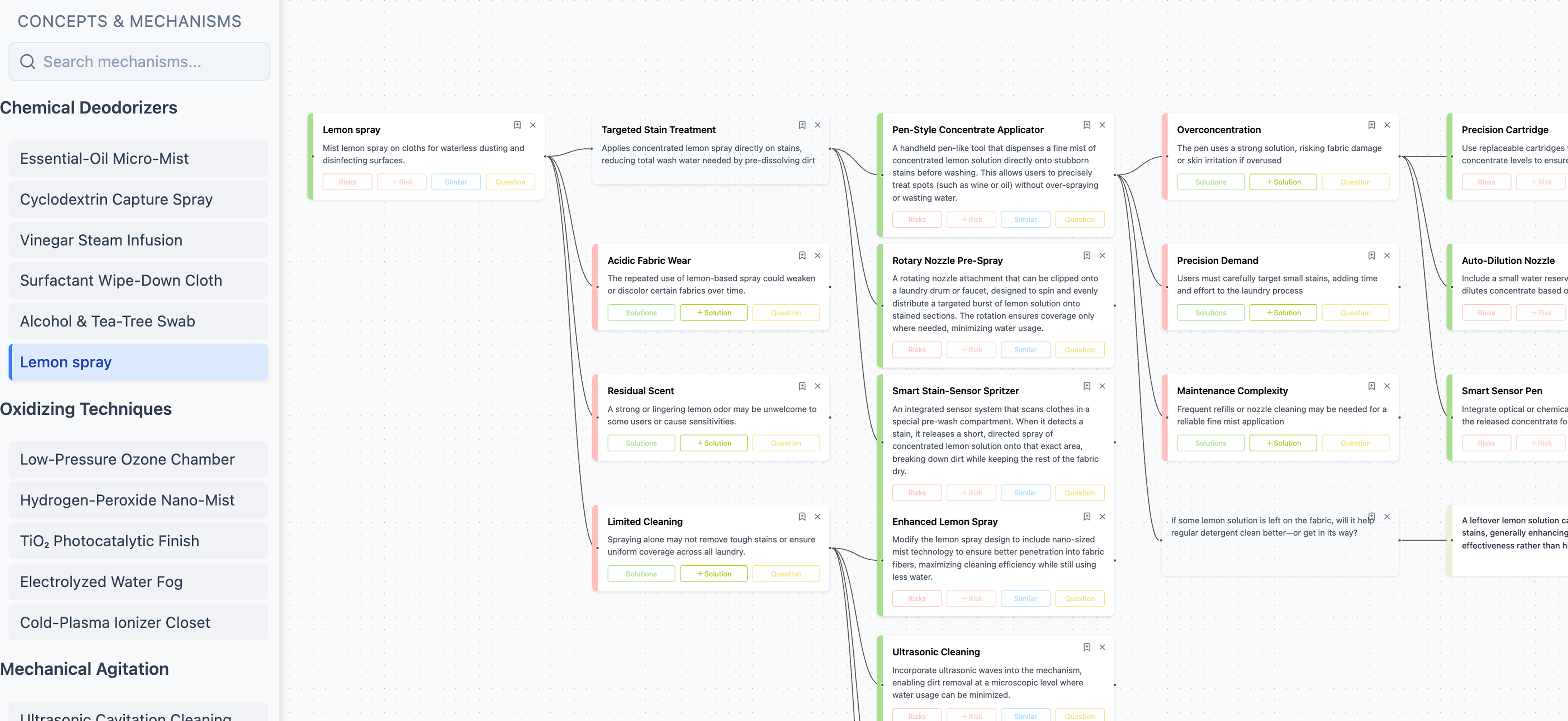}
  \caption{ User and AI collaboratively work on selected ideas in canvas view.}
  \Description{User and AI collaboratively work on selected ideas in canvas view.}
  \label{fig:overview-d}
\end{figure}


We instantiated these supports in a prototype system, \sys, where we use the OpenAI API to access the GPT-o1 model on the backend. We use the following use case to demonstrate the workflow. 

Consider an R\&D expert, Sarah, who is brainstorming on the design challenge: `how to clean laundry with less water'. Struggling to generate ideas from scratch, she turns to \sys and inputs the task. The system prompts a LLM with a structured query designed to elicit diverse idea categories, each accompanied by concrete examples. It returns a set of categorized ideas, which Sarah explores by hovering over category and idea names to view detailed descriptions(Figure~\ref{fig:overview-b}). 
Among the results, Sarah is drawn to the `Chemical Deodorizers' category and selects the `Lemon Spray' idea, which opens a canvas page(Figure~\ref{fig:overview-d}) for further investigation.

\begin{wrapfigure}{r}{0.45\textwidth} 
\centering
\includegraphics[width=\linewidth]{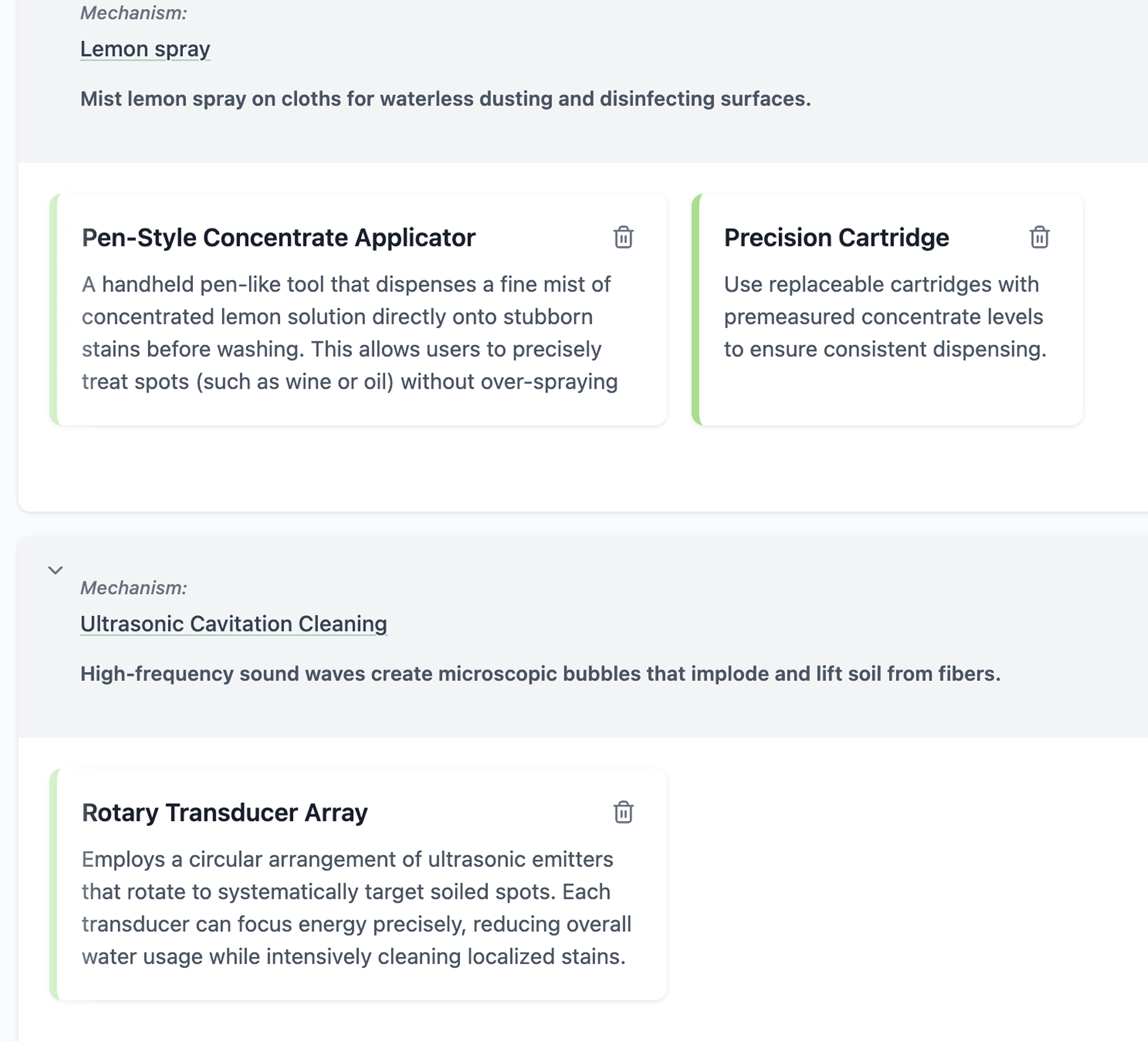}
\caption{ User pins and further develops on the promising ideation outcomes.}
\Description{User pins and further develops on the promising ideation outcomes.}
\label{fig:overview-e}
\end{wrapfigure}

 Wanting to discover similar ideas to this seed idea, she clicks the similar button (Figure~\ref{fig:components}a). The system then presents  related idea categories based on the abstract concepts underlying lemon spray that enable its cleaning effect. In addition to the original category `chemical deodorizers', the system also identifies and presents her new categories like `Targeted Stain Treatment' and `Natural Acidic Boost Appear'. This reflection-oriented process helps Sarah clarify what she values in the idea, enabling the system to offer suggestions better aligned with her preferences rather than relying on the LLM’s randomness to surface relevant alternatives. 

Realizing she is drawn to the targeted cleaning aspect of lemon spray, Sarah clicks on `targeted stain treatment' (Figure~\ref{fig:components}a) to explore more ideas. She finds an interesting idea,  `pen-style concentrate applicator', and saves it for future use by clicking the collection icon.
The idea also sparks a question: If some lemon solution remains on the fabric, will it enhance or interfere with detergent? She clicks the `question' button to input the question, and the system incorporates her query into a prompt for the LLM. It then returns a concise, context-aware answer (Figure~\ref{fig:components}c). 

To further evaluate the original `lemon spray' idea, Sarah returns to its card  and clicks the `risk' button (Figure~\ref{fig:components}b) to review potential drawbacks. She identifies `limited cleaning' as a key tradeoff and clicks the solution button on the corresponding card to explore mitigation strategies. While reviewing one solution: add an additional cleaning agent to complement the lemon spray, Sarah realizes she could substitute lemon juice with a mild oxidizing agent while using lemon essential oil to retain a pleasant scent. She then adds her own solution: a hydrogen peroxide and lemon mix spray (Figure~\ref{fig:components}d).

The sidebar continuously updates with both initial and newly generated idea categories and ideas. Using the sidebar (Figure~\ref{fig:overview-d}), Sarah navigates between concepts, explores promising directions, and collects those that interest her. She then clicks on the `explored canvas' tab to view a summary of all the ideas she has collected (Figure~\ref{fig:overview-e}). From there, she can continue brainstorming within the system or move selected ideas forward into further development or prototyping.

\section{Future work}
We conducted an initial test with three users to compare two conditions. In the baseline condition, users completed the ideation task `minimizing accidents from people  walking and texting on a cell phone'~\cite{ma2023conceptual} using any available information sources, including ChatGPT and a browser. In the experimental condition, users completed the task `cleaning laundry with less water' \footnote{https://www.mindsumo.com/contests/dry-laundry} using our system, while having access to the browser and ChatGPT. Users reported \sys 
helped them quickly analyze risks and identify alternative ideas they were interested in. Compared to the baseline, the system provided three useful forms of support and actionable cues without limiting flexibility. This flexibility came from allowing users to control the sequence of actions, revisit prior steps, and contribute their own thoughts, which the AI used as contextual input.

These results suggest \sys has the potential to enhance human–AI collaborative ideation for creative problem-solving. However, much work remains to fully realize this potential. More user tests and comprehensive evaluations are currently underway. We plan to measure the quantity, quality and users' experiences with the system, and use this feedback to improve the prototype for a smoother human–AI co-ideation process. 

\begin{acks}
This work was supported by the Toyota Research Institute, the Office of Naval Research, and partially supported by the National Science Foundation under Grant No. \#2118924 Supporting Designers in Learning to Co-create with AI for Complex Computational Design Tasks.
\end{acks}

\bibliographystyle{plainnat}
\bibliography{sample-base}

\ccsdesc[500]{Human-centered computing~Interactive systems and tools}

\end{document}